\documentclass[12pt]{article}

\usepackage{graphics,epsfig}

\usepackage{amssymb}

\begin{document}
\title{Global Physics:\\ From  Percolation to
Terrorism, \\Guerilla Warfare and Clandestine Activities}

\author{Serge Galam\\
 Laboratoire des Milieux D\'esordonn\'es et
H\'et\'{e}rog\`enes,\\ Tour 13 - Case 86,
4 place Jussieu, \\ 75252 Paris Cedex 05}
\date{(galam@ccr.jussieu.fr)}
\maketitle

\begin{abstract}
The September 11 attack on the US has revealed an unprecedented terrorism with worldwide range 
of destruction. It is argued to result from the first worldwide percolation of 
passive supporters. They are people sympathetic to the terrorism cause but without
being involved with it. They just don't oppose it in case they could. This scheme puts
suppression of the percolation as the major strategic issue in the fight
against terrorism. Acting on the population is shown to be useless. Instead a new strategic
scheme is suggested to increase the terrorism percolation threshold and in turn suppress the
percolation. The relevant associated space is identified as a multi-dimensional social space
including both the ground earth surface and all various independent flags displayed by
the terrorist group. Some hints are
given on how to shrink the geographical spreading of terrorism threat. The model apply to a
large spectrum of clandestine activities including guerilla warfare as well as tax evasion,
corruption, illegal gambling, illegal prostitution and black markets.

\end{abstract}

{keyword:}
terrorism,  percolation,  passive supporters, social space\\


\section{Defining the problem}

Terrorism is a daily reality of horror for millions people in the world. Though it has 
always existed, its various expressions have been confined within quite precise
areas. However, September 11 has revealed an unprecedented world wide spreading of a given
terrorist group. In response to such a new and permanent world threat, enormous 
military, police and information efforts are put in to try to curb this new terrorism. But
in spite of a certain number of significant but always specific
successes, the current terrorist challenge remains incredibly powerful and present. 

It is worth to note, though at a different scale, the same holds true for more traditional
terrorism like Basque, Irish and Corsican ones. After years of anti-terrrorism fight
they are still not eradicated quite to the contrary. It thus seems that the
techniques used to fight terrorism are not adapted to the task. We  suggest here focusing onto
the social space in which the terrorists live, moves and acts. While much effort are
devoted to study terrorist networks themselves, their structure, their means, their
motivation, and their potential targets \cite{loup,sandler}, very little is known about the
human environment in which terrorists evolve. This space includes the terrorists themselves,
their potential targets and also each one of us. 

The following of the paper is organised as follows. Part two introduces the notion of
social permeability to terrorism. The problem of the range of terrorism threat is then
analysed in terms of a percolation phenomenon. Traditional terrorism corresponds to non
percolating situations while international terrorism is associated to a world-wide
percolation. Curbing terrorism using only military means is found to be ineffecient in Part
three. Part four suggests a new strategy to reduce current world-wide terrorism
threat without hurting its passive supporters. Some hint on how to shrink the terrorism threat
without massive military destruction are singled out. The last part discuss the universality
of the model. In particular it is applicable to guerilla warfare as well as tax evasion,
corruption, illegal gambling, illegal prostitution and black markets.

\section{The social permeability}

We study the social permeability to terrorist individual moves. Permeability is used here
to mean a free passage granted to a terrorist. It is assumed to result from passive
support to the terrorist cause by some subset of the population. It holds
without any involvement with either the terrorist group or its activities. It is an
individual dormant attitude associated to a personal opinion. It does not need to be
explicitly set. We denote these people as passive supporters. They just do not
oppose a terrorist act in case they could (see Fig. (1)).  They go unnoticeable and 
most of them reject the violent aspect of the terrorist action. They only share
in part their cause.

\begin{figure}
\begin{center}
\centerline{\epsfbox{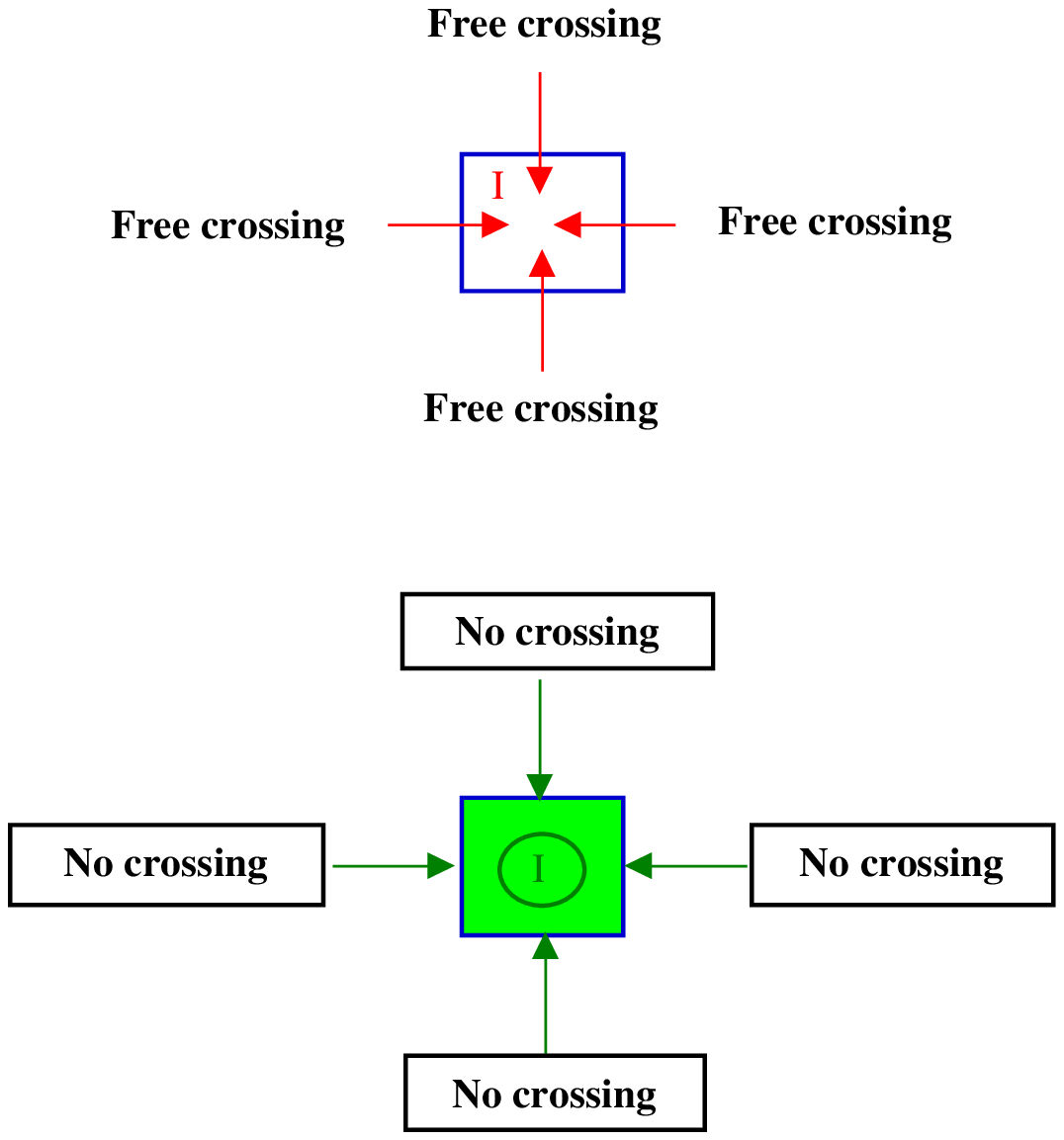}}
\caption{On top a box in passive position, "closed curtains", a terrorist can cross the
box from any of its four sides. On bottom, a box in active position, "opened
curtains", a terrorist cannot cross safely. From any side the box is under scrutiny.}
\end{center}
\end{figure}    

We are dealing here with a distribution of individual passivities in sharing 
independently a common sympathy towards the terrorist cause. Passive supporters do not
need to communicate among them. Within a metaphorical representation, it can be said that
each individual is looking only out of its own window, and not beyond. In this schematic
way, each person occupies and observes some portion of a territory from its personal window.
Each one can then decide to close or open
its window curtains independently from each other at the time a suspect observation could
be done. 

Accordingly, to move freely and safely a terrorist must find a series of
contiguous windows of closed curtains. To be reachable from the terrorist base a potential
target must be connected by at least one continuous path of closed curtain windows.
Since indeed each window state is independent of the terrorist will, it is in fact all of
the existing possible paths starting from the terrorist base, which determine the social
space open to terrorist action. We call it the Active Open Space (AOS). Several 
Open Spaces (OS) are expected to exist simultaneously but are not accessible to
terrorist action, not being connected to the terrorist base. An illustration is given in Fig.
(2)

\begin{figure}
\begin{center}
\centerline{\epsfbox{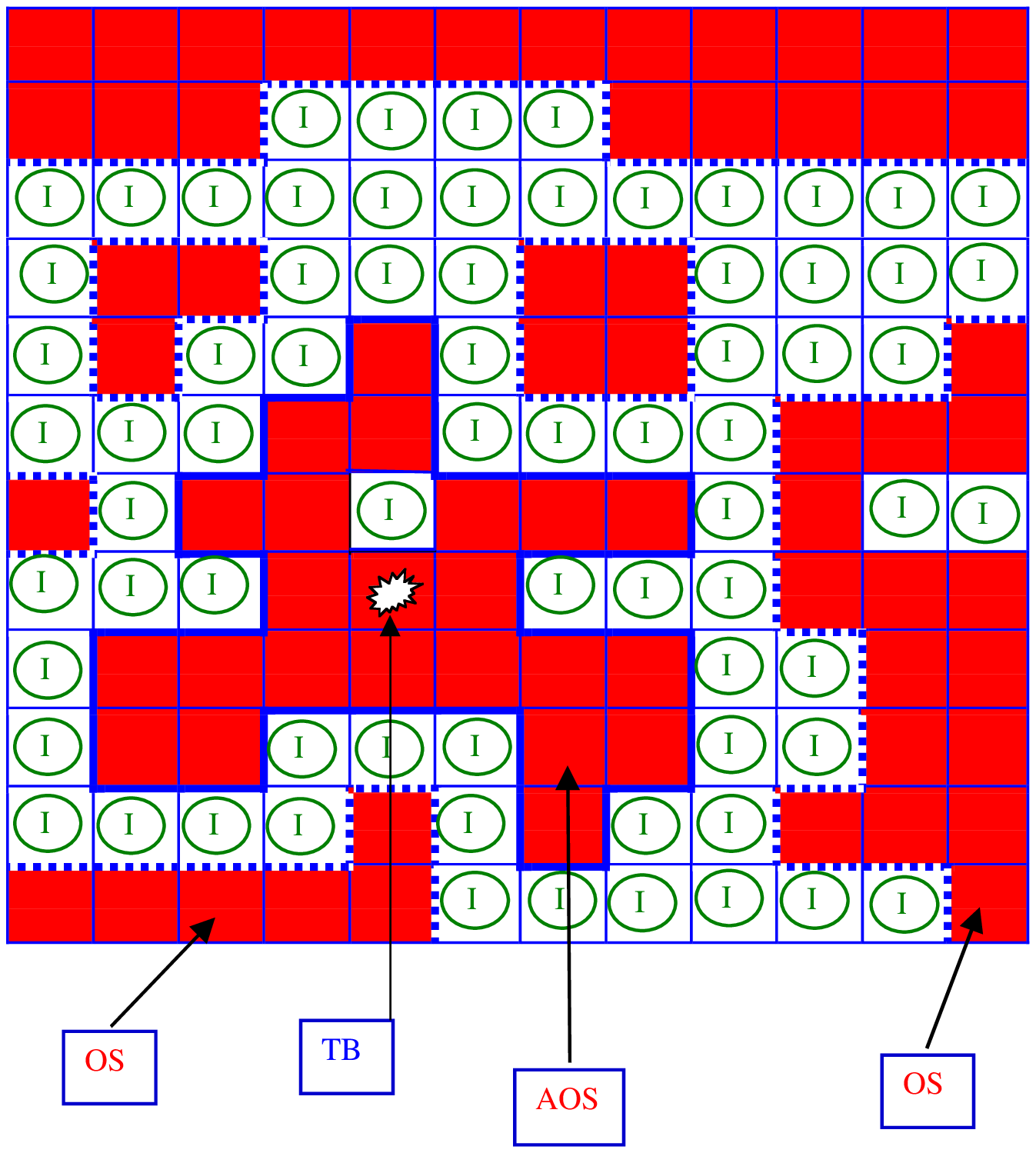}}
\caption{A portion of 12x12 box grid. Passive supporters occupy dark boxes, their 
curtains are closed. The boxes are free to terrorist crossing. On the other hand, white
boxes have open curtains, they are closed to terrorist movement. The 
nearest neighbor juxtaposition  of dark boxes produces Open Spaces (OS) to terrorist
action. But only the one including the terrorist base (TB) is active (AOS) with an
effective terrorist threat. Other OS are inaccessible to terrorist action.}
\end{center}
\end{figure}    

It is worth to note the huge majority of passive supporters will never be
faced with  a terrorist crossing by. These individual sympathies are dormant and almost
never activated. It makes them invisible and non-identifiable within an anti-terrorist
fight. They are completely scattered and randomly distributed on a given territory.
Nevertheless it is their geometrical aggregation by simple adjacent juxtaposition, which
creates a series of labyrinths of permeable paths to terrorist moves. 
Not withstanding their invisibility they determine the whole spectrum of targets
potentially accessible to a terrorist action. In return any attempt to circumvent the
terrorist threat cannot be successfully achieved without an evaluation of the current
degree of social permeability. It is indeed both a strategic and a conceptual challenge.
And here is where Physics can proves useful.

At this stage, applying percolation theory to terrorism social permeability appears quite
natural in order to build a coherent and unified framework to the geographical deployment of
terrorist action
\cite{terro1}. An illustration is exhibited in Fig. (3) Such a "Global Physics" approach
\cite{first} is part of a new growing trend of research
\cite{sex}. In the last years a growing number of physicists have been studying 
social and political behaviors using concepts and tools from Statistical Physics
\cite{monde,frank,helb,nadal,Krause,poli}. Among others, a rumor formation model has been
used recently to explain the French Pentagon Hoax case \cite{hoax}. 
Percolation theory  \cite{perco} has been also used to describe connectivity problems
\cite{pajot,perco-sor}.  Here it allows to connect the range of a given terrorism threat to
the surrounding population attitude towards that specific terrorism.
\cite{terro1}. However it is worth stressing we are are not investigating either the terrorist
net itself or its internal mechanisms.

Moreover our work does not aim at an exact description of terrorism complexity. Making some
crude  approximations allow exhibiting an essential characteristics of terrorism by
linking its capacity of destruction to the surrounding population attitude. In particular
a target is set to be reachable once it is located within an area covered by a cluster of
people who are passively consenting to the terrorist cause. The September 11 terrorist
attack on the US is given an explanation in terms of the first worldwide percolation of
such a cluster of passively consenting people \cite{terro1}. On this basis some clues are
obtained on how to curb terrorism threat without using large range military destruction
\cite{terro2}.

\section{The no solution scenario }

At this stage, we can conclude that for a given territory the distribution and the size 
of passive supporter aggregated spaces yield the range of terrorist action. It is the
relative value of the passive support $p$ of the population compared to the value of the
critical threshold $p_c$ of the corresponding space, which determines the effective
amplitude of the terrorist threat. If the passive support density is less than the percolation
threshold ($p<p_c$), most of the territory is safe with only one rstricted area under
terrorist threat as shown in Fig. (3). At contrast as soon as the passive support density gets
larger than the percolation
threshold ($p>p_c$), all the territory falls under the terrorist threat (see Fig. (4)).

\begin{figure}
\begin{center}
\centerline{\epsfbox{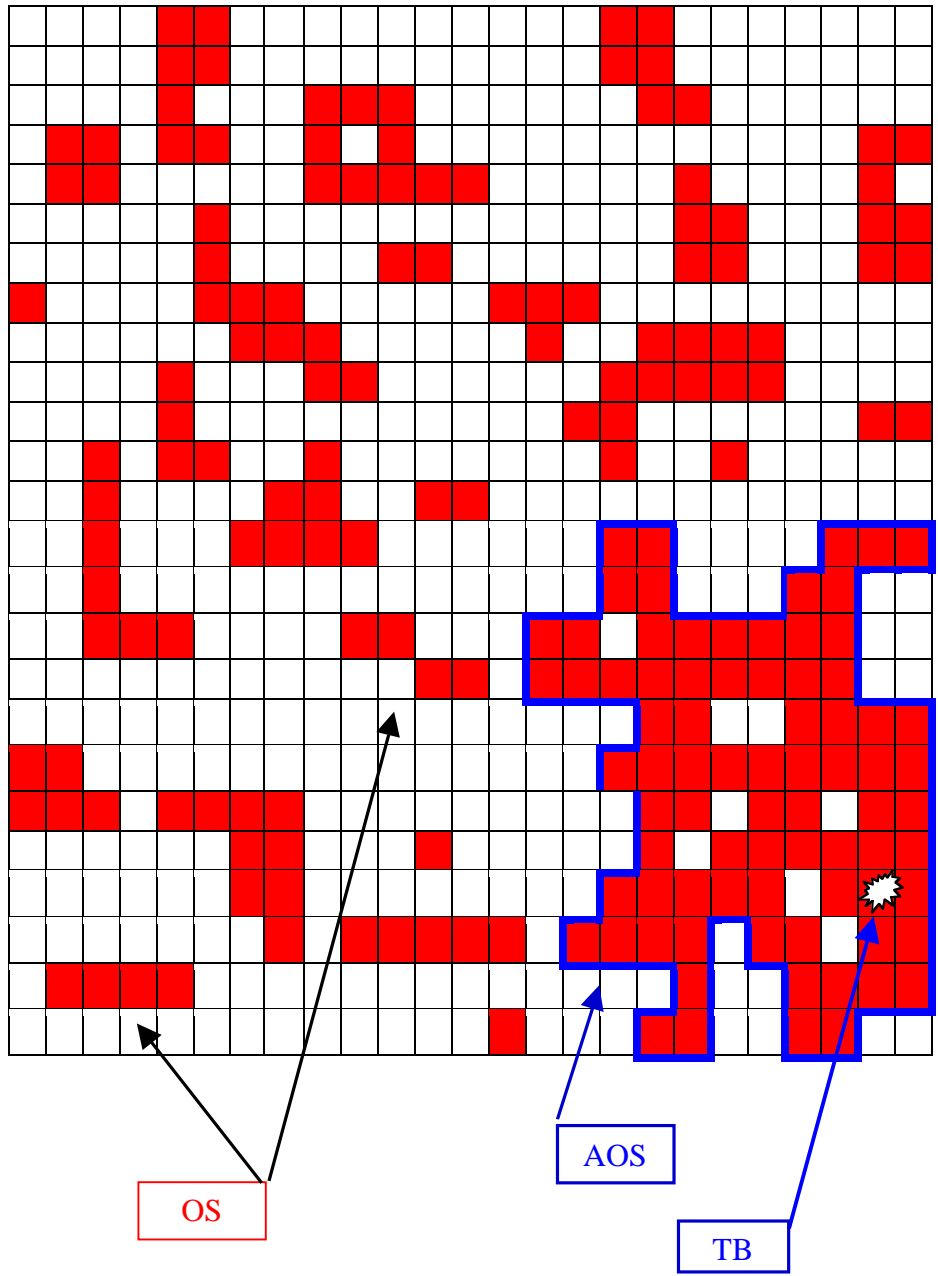}}
\caption{A whole country sparse with little and disjoined OS of more or less wide. They are
indicated in dark. The couintry is not under total terror threat.}
\end{center}
\end{figure}    

\begin{figure}
\begin{center}
\centerline{\epsfbox{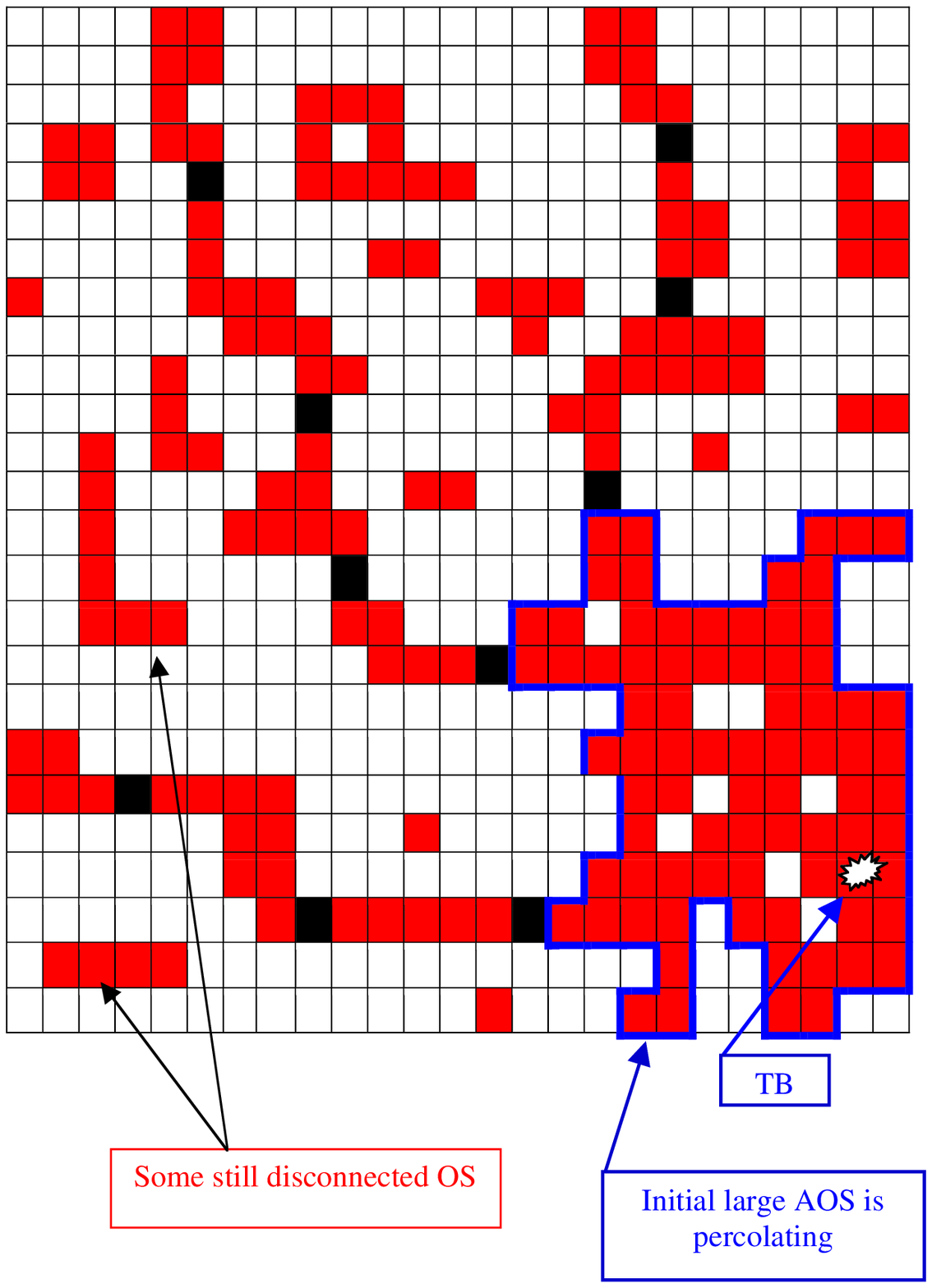}}
\caption{Few additional passive supporters have appeared with respect to Fig (3). They are
sufficient to create a percolation phenomenon. The whole country is now under total
terror threat (shown in dark). Some sparse OS are still disjoined from the percolating
cluster.}
\end{center}
\end{figure}

However, while in a physical system the size is infinite, here the
delimitation of the size of the territory considered is an essential data. One can have a
density of passive supporters $p$ which satisfies the condition $p>p_c$ over a given
geographical surface, and at the same time for a larger territory including the first one
another density $pÕ$ satisfying $p'<p_c$. Indeed, always the number of passive supporters
strongly decreases far from the home area of terrorism. That is the case of traditional
terrorism already mentioned previously. For example in the Corsican case,
terrorism does percolate at the level of the associated island but not at the level of
France and even less at the world level.

Terrorist deployment thus obeys a universal scheme of activity with two phases, a 
percolating one and a non-percolating one. The only difference from one terrorism to
another is the scale on which passive supporters are spread and the geographical
area on which a percolation may take place. Obviously if the change of scale does not
change the nature of the terrorist phenomenon, it modifies in a substantial manner the
number of threatened people. It is clearly not a negligible difference. 
It is the case of September 11, which while revealing for the first time the existence 
of a world percolation showed that simultaneously from now on the whole world population
is in danger.

In this context, for a territory under active terrorist threat, the lifting of the 
threat requires the suppression of the  passive supporter percolation. If
any destruction of a terrorist cell has obvious immediate advantages, it is without effect
on the range and acuity of the threat. As soon as a new terrorist group is formed, it can
strike again and immediately in all the space which remains accessible to its members. The
strategic challenge is thus to bring back the condition $p>p_c$ to some new
condition $pÕ<p_c$ by lowering the density of passive supporters from $p$ to $pÕ$. Such a
reduction induces a sudden shrinking of the whole territory accessible to terrorist action.
Such a condition would reduce the AOS to a narrow geographical area. However a
solid implementation of such a programm on the military level is simply terrifying.

Therefore an efficient military solution is completely unacceptable for ethical reason as 
well as from the point of view of moral and justice. It feasibility would lead destroying
a good part of the planet, although existing weapons of massive destruction would allow
it. At the same time, any partial military solution appear to be useless since
without effect on the level of the terrorist threat. At this stage the
conclusion from our study seems to show no hope in curbing terrorism. Accordingly,
international terrorism would be a fatality against which nothing can be done. The
anti-terrorist fight being limited to specific actions against the networks themselves.
The current danger would thus remain unchanged with a world social permeability activated
on request from a terrorist group.

\section{The social space of terrorism}

From above no solution conclusion, reversing the usual physical scheme to suppress a
percolation can proves usefull in providing an alternative to the problem. Instead of
changing the density of sites,  a solution may
exist on modifying the value of the percolation threshold itself without touching on the sites
themselves.

Percolation thresholds 
depends primarily  on two independent parameters, the connectivity of the network $q$ and
the dimension of the space $d$. For instance, for a square lattice with
$q=4$ and $d=2$, $p_c=0.59$. On the other
hand the cubic extension has $d=3$,  $q=6$ and $p_c=0.31$. The
hypercube at four dimensions has $q=8$ and $p_c= 0.20$. 
Increasing the dimension or and the connectivity 
creates more possible ways to connect from one site away. It leads to a reduction in the
value of the percolation threshold. However very few geometrical networks allow an exact
calculation of their percolation threshold, most of them being calculated numerically. 

For a social application of percolation, connectivity may be of the order 
of 16 for a dimension of a priori 2, the surface of the earth. It
would correspond to an unknown network in Physics. Thus its percolation threshold
is unknown. But fortunately it turns out that few years ago a universal formula for all
percolation thresholds was discovered \cite{gama}.  

\begin{equation}
p_c = a[(d-1)(q-1)]^{-b}\ ,
\end{equation}
where $d$ is dimension, $q$ connectivity, $a= 1.2868$ and $b = 0.6160$. The Galam-Mauger
formula \cite{gama} yields with the percent close all known thresholds. In addition it can
predict the value of the threshold for any network defined by the values of both its
connectivity and dimension. The formula is shown in three dimensions in Figure (5).

\begin{figure}
\begin{center}
\centerline{\epsfbox{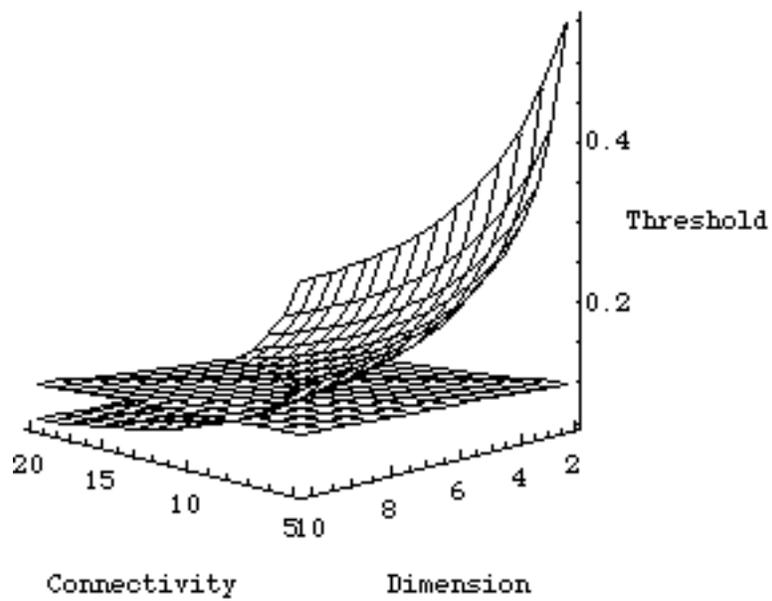}}
\caption{Representation of the  Galam-Mauger universal formula \cite{gama} for all the
thresholds of percolation as function of connectivity and dimension. The formula writes $p_c
= a[(d-1) (q-1)]^{-b}$ where $d$ is dimension, $q$ connectivity, $a= 1.2868$ and $b = 0.6160$.}
\end{center}
\end{figure}    

Having the formula, assuming $q=16$ the challenge is to estimate the dimension of the
terrorism social space. Here we make the hypothesis that in a social movement additional
dimensions are produced by social paradigms in which individuals may position themselves
in a similar way they do on earth. Thus it is necessary to consider the "flags" by which
terrorists ground their fight in addition to the earth surface two dimensions \cite{terro2}. To
each one of these flags, people may identify with more or less support.

Typically for most terrorist groups, the first flag is a territorial claim
either independence or autonomy.  This flag constitute the first social dimension
independent of those from the geography. Also, as soon as a terrorist group enters in
action, it induces some state repression against it, which in turn determines a new
additional flag. People may disapprove the repression hardness. That gives already 4
dimensions. Thus, any terrorist social dimension seems to be at least of a value four.

For $q=16$ and $d=4$ the Galam-Mauger formula \cite{gama} yields a percolation
threshold at $p_c=0.12$ to be compared with the value $p_c=0.65$ for a square system with
$q=4$ and $d=2$. It means that as soon as a terrorist causes is supported by
more than $12\%$ of a population, the corresponding terrorist group can move with
complete freedom on all the associated territory. This value is not very high making it
certainly reached in traditional terrorism in Corsica, North Ireland and Basque area (see Fig.
(6). It may explain the continuous on going incapacity of the respective authorities to cut
short these terrorist groups. 

\begin{figure}
\begin{center}
\centerline{\epsfbox{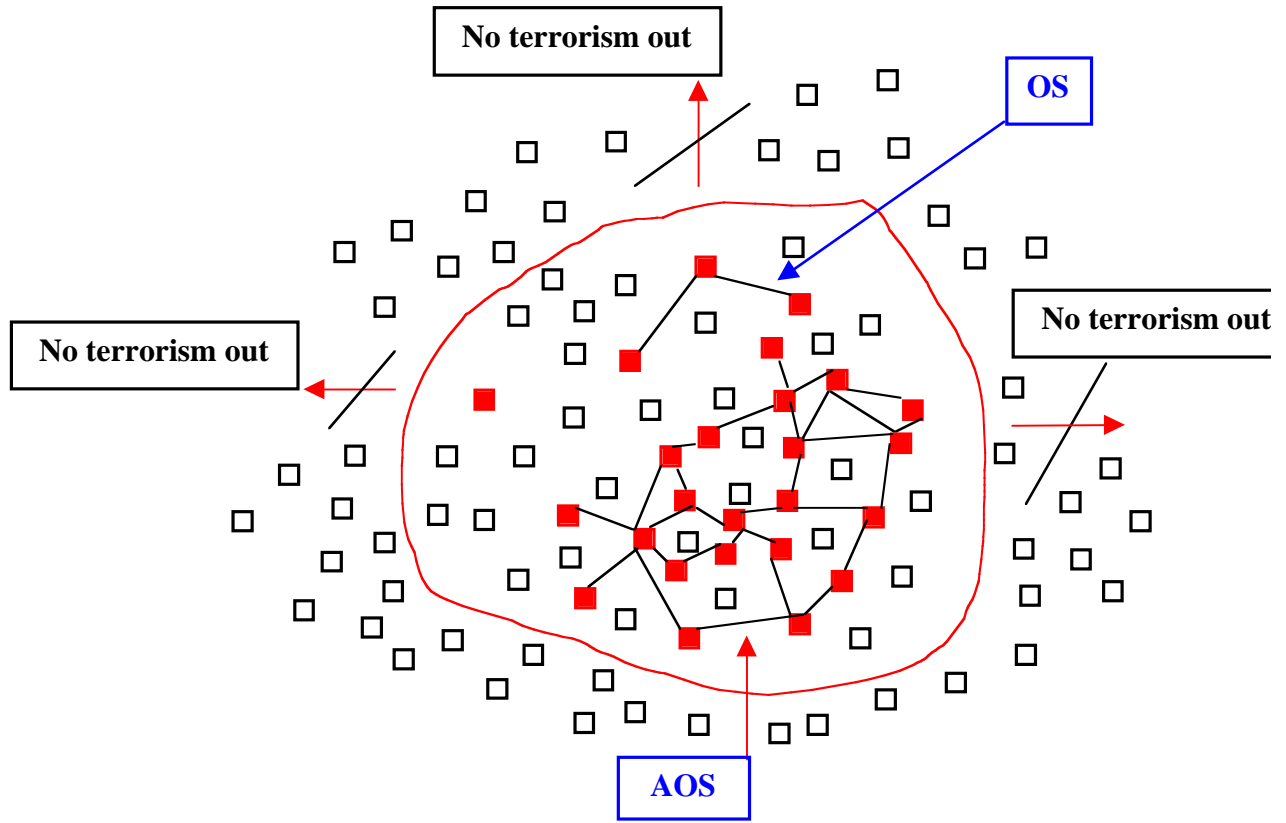}}
\caption{Diagram for Corsica with an Active Open Space, which percolates at the level of 
the island but with no possible extension beyond. Black squares are passive supporters while
while ones are not.}
\end{center}
\end{figure}

In above cases, within the framework of our model, not much seem to be possible 
since a dimension of 4 is irreducible. It is the dimensional lower limit of
any terrorist activity. Traditional terrorism of low dimension could thus keep alive,
except finding how to yield a drop in the density of passive supporters down to less than
$12\%$. It is interesting, and perhaps useful to note that, within the present framework 
the absence of repression would bring down the dimension at $d=3$ making the threshold up
to  $16\%$. Such numbers would suggest that for instance, in case the popular Corsican
support for independence ranges between $12\%$ and $16\%$, it is indeed the repression
which allows it to percolate all over the Corsican territory.

With regard to the new international terrorism the situation seems 
qualitatively and quantitatively different. Indeed it is difficult to believe it has a
support of more than $10\%$ of the world population. At the same time it appears to be
clearly successful in having its passive supporters to percolate worldly. Using the
Galam-Mauger formula \cite{gama}, to carry out a percolation with only a few percent of people
and a connectivity of 16 requires a space of higher dimension larger than the 4 value of
traditional terrorism as shown in Fig. (7).

\begin{figure}
\begin{center}
\centerline{\epsfbox{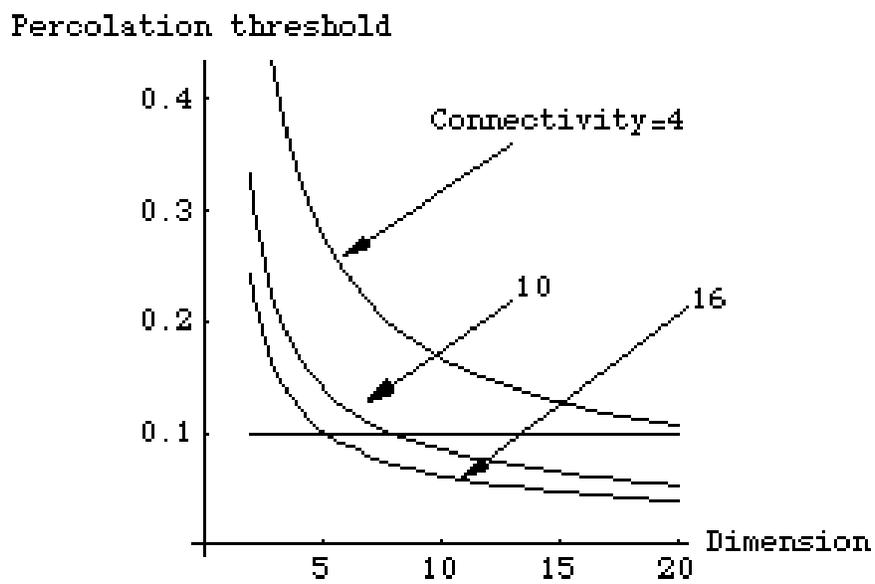}}
\caption{Representation of the universal formula Galam-Mauger \cite{gama} for fixed
connectivity as function of dimension. It is seen that the threshold values drop with
dimension}
\end{center}
\end{figure}

And indeed, what characterizes current international terrorism 
is the broad spectrum of flags on which it deploys its claims \cite{terro2}. In addition to 
traditional territorial claims, it has at least a religious dimension, an ethnic dimension,
a bipolarizing dimension of partitioning the world, a social dimension, a regional
dimension and a historical dimension. That brings its social dimension to 10.
Such a value results in a lowering of the percolation threshold down to only $6\%$, which
becomes a more realistic estimate for its world support. 

\section{Hints for global terror curb}

Current world terrorism by creating a large number of independent flags has drastically
increased its potential range of world destruction. That process has enabled 
a world percolation of only few percent of passive supporters all over the world.
But simultaneously, it provides hints to envision some solution in curbing current
world threat. In particular, contrary to low dimension terrorism for which a dimension
reduction is impossible, here action becomes possible to reduce the number of
dimensions deployed by international terrorism \cite{terro2}. More specifically, political,
economic and psychological actions should be capable to neutralize a certain number of these
flags without use of military means.

For example to lower the social dimension from $d=10$ to $d=6$ make the threshold to 
soar
from  $6\%$ to $10\%$. Such an increase would suppress at once the world percolation.
Accordingly terrorism would immediately reduce to only one area of the world as for 
other
terrorism. Of course the solid measures on how to put in action
the flag neutralization process is out the scope of physicists. It requires an
interdisciplinary collaboration with specialists in the other concerned disciplines.

At this stage the possibility of an efficient non-destructive large scale fight
against international terrorism passes trough a deep interdisciplinary research. What
may turn a challenge may be more difficult than the fight against terrorism. Last but
not least it is worth to restate the current analysis does not claim at an exact
quantitative description of terrorism reality. It only aims in shedding a new light on
linking passive individual support to a terrorism cause and the associated range of
action of the corresponding terrorist group.

\section{From terrorism to guerilla warfare and undeground activities}

Our model to terror threat is indeed much more universal than the field of terrorism.
First we have to define more precisely what terrorism is. In particular what is denoted
terrorism by legal and institutional organizations is called freedom fights by others
underground groups. Most of the time, terrorism is what the others do. Nevertheless some
consensus seems to exist  along for instance the so called ``Mitchell report", which was
accepted by both the Palestinian Authority and the Israeli Government, defining
terror as attack on random, unarmed people. 

While our model definitively apply for above definitio, it also applies equally well to the
murder of a specific target by a clandestine group, like the killing of Israel's tourism
minister Ze'evi, which is not terrorism in the Mitchell sense. Accrodingly guerilla warfare
may be a more appropriate word for violent activities modeled by the model.

Moreover, the same mathematics also applies to lots of other clandestine activities like tax
evasion, corruption, illegal gambling, illegal prostitution, black
markets, etc. Without undocumented immigrants, the economy of California might break
down. On this basis we
can conclude that we have presented indeed a universal model to clandestine cooperation.
It would be now very fruitfull to confront our general frame to real cases with data and facts
But clearly such a taskis beyong the physicist skill.

\end{document}